# Long Jump

V. Dorobanţu

Physics Department, Timişoara "Politehnica" University

*Abstract.* When the Physics' laws are taken seriously, the sports can benefit in getting better results, as was the case of the high jump in Fosbury flop style: sprinting diagonally towards the bar, then curve and leap backwards over it. The jumper's centre of mass being under the bar, this style allows improvement of the performance.

**Theory**

Considering $\vec{v}$ (t), the velocity of the sportsman's centre of mass, and **F**, the force acting upon him, we write, according to the Newton's second law [1]:

$$\partial_t(m\vec{v}) = \vec{F} \qquad (1)$$

with

$$\vec{F} = \vec{R} + \vec{G} \qquad (2)$$

$\vec{R}$ is the friction force with the air and $\vec{G}$ the weight of the athlete.

The resistance force, for a body moving into a fluid inspired by the Bernoulli's dynamic pressure, can be taken as:

$$R = A\frac{\rho v^2}{2} c_d \qquad (3)$$

$c_d$ being the drag coefficient [2], A, the normal area exposed to the fluid, v the speed of the athlete and $\rho$ the fluid's density. The speeds involved in such kind of sports are small enough, so the mass remaining constant and splitting the motion into Ox and Oy directions, we have:

$$m\,\partial_t(\partial_t x) = -c_{dx}\frac{\rho A_x}{2}(\partial_t x)^2 \qquad (4)$$

$$m\,\partial_t(\partial_t y) = -c_{dy}\frac{\rho * A_y}{2} * (\partial_t y)^2 - mg \qquad (5)$$

or

$$m\,\partial_t v_x = -c_{dx}\frac{\rho * A_x}{2} * v_x^2 \qquad (6)$$

$$m\,\partial_t v_y = -c_{dy}\frac{\rho * A_y}{2} * v_y^2 - mg \qquad (7)$$

Using **Mathematica** to solve the equations of motion, from (6) and (7), we get:

$$v_x = \frac{v_0 \cos\alpha}{1 + \frac{A_x \varrho\, t\, c_{dx}\, v_0 \cos\alpha}{2m}} \qquad (8)$$

$$v_y = (\sqrt{2gm}\ \mathrm{tg}[\arctan[(v_0\sqrt{\varrho A_y c_{dy}}\sin\alpha)/(\sqrt{2gm})] - \tfrac{t\sqrt{g\varrho A_y c_{dy}}}{\sqrt{2m}}])/(\sqrt{\varrho A_y c_{dy}}) \qquad (9)$$

Knowing that $v_x = \frac{dx}{dt}$ and $v_y = \frac{dy}{dt}$, integrating equations (8) and (9) we have:

$$x = x_0 - \frac{2m \ln(2m)}{A_x \varrho\, c_{dx}} + \frac{2m \ln(2m + A_x t \varrho\, c_{dx}\, v_0 \cos\alpha)}{A_x \varrho\, c_{dx}} \qquad (10)$$

$$y = y_0 + \frac{m \ln[1 + \frac{\varrho (\sin\alpha)^2 A_y c_{dy} v_0^2}{2gm}]}{\varrho A_y c_{dy}} + \frac{1}{\varrho A_y c_{dy}} 2m \ln[\cos[\arctan[(v_0\sqrt{\varrho c_{dy} A_y}\sin\alpha)/(\sqrt{2gm})] - \frac{t\sqrt{\varrho g A_y c_{dy}}}{\sqrt{2m}}]] \qquad (11)$$

When the areas and/or drag coefficients are zero, one gets the results for a body moving without friction:
$v_x(A_x \to 0) = v_0 \cos\alpha$, $x\ (A_x \to 0) = x_0 + v_0 t \cos\alpha$, $v_y(A_y \to 0) = v_0 \sin\alpha - gt$, $y\,(A_y \to 0) = y_0 + v_0 t \sin\alpha - \frac{gt^2}{2}$.

Since we are interested in how long can jump the athlete, eliminating time, t, between eq. (10) and (11) we get x(y, α):

$$x = \frac{1}{\varrho A_x c_{dx}}(2m(-\ln(2m) + \ln[(\sqrt{m}(2\sqrt{gm A_y c_{dy}} + \sqrt{2\varrho}(\arccos[(e^{\frac{\varrho A_y c_{dy}(y-y_0)}{2m}} gm\sqrt{4 + \frac{2\varrho(\sin\alpha)^2 A_y c_{dy} v_0^2}{gm}})/$$

$$(2gm + \varrho(\sin\alpha)^2 A_y c_{dy} v_0^2)] + \arctan[\tfrac{v_0\sqrt{\varrho A_y c_{dy}}\sin\alpha}{\sqrt{2gm}}])(\cos\alpha) A_x c_{dx} v_0))/(\sqrt{gA_y c_{dy}})]) + \varrho A_x c_{dx} x_0) \qquad (12)$$

In order to get the maximum horizontal length, we have to solve:

$$\partial_\alpha x = 0 \qquad (13)$$

The solution will give us the appropriate inclination angle with the horizontal plane in order to get the maximum length.

**Numeric results**

As one can be seen from the above equations, the maximum horizontal length of the jump depends of a lot factors: the mass of the jumper, the drag coefficients (horizontally and vertically), $y_0$ the height of the jumper's mass centre, $A_x$ the normal area exposed to the fluid in horizontal moving, $A_y$ the normal area exposed to the fluid in vertical moving, air density, the gravitational acceleration and the initial velocity.

Taking: $y_0$ =1.06 m (according to the Da Vinci's Vitruvian man[3], the center of mass is placed at 61.8% of the height, if standing and, at ran, it is reduced by 1/14 from his height ), g=9.8 m/s², ρ=1.293 Kg/m³, m=94 Kg, $v_0$=10.44 m/s, $c_{dx}$=1.1 (similar to that of the skier [2]), $c_{dy}$=0.5, $A_x$=0.77 m² (an approximation of the frontal area of the jumper), $A_y$=0.25 m² (an approximation of the head an shoulders' area), $x_0$=0, y=0.21 m (when the sportsman lands, his mass center is 10.8% from his height above the ground), and solving the equation (13) we get **α=42⁰ 54 minutes**. So, the sportsman having the above data must jump at an angle of **42⁰ 54 minutes,** with the horizontal plane, and an initial velocity of 10.44 m/s, in order to reach **11.54 m.** This could be the length, if Usain Bolt ( his data have been taken into account [4]), would jump.

Here is graph showing the length of the jump as a function of the initial velocity and the inclination angle, for a sportsman with characteristics mentioned above.

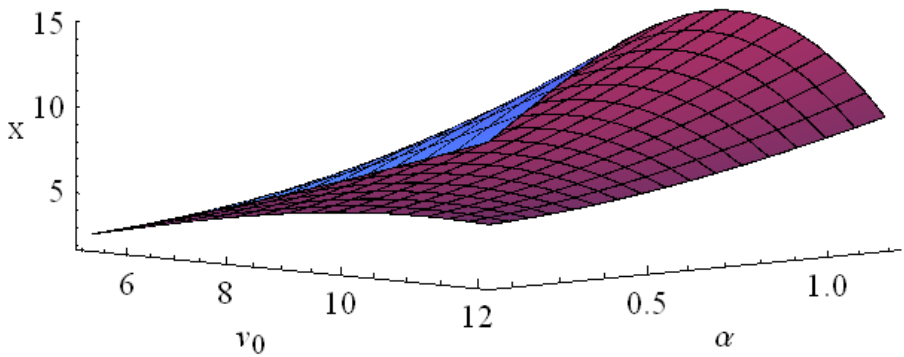

Fig.1

**Conclusions**
The above calculations must be adapted for every sportsman, in order to get maximum from his/her qualities and, also, the place where the jump is done plays some role. For instance, at Mexico City (2240 m, high) where the density of air is reduced by 27% and at an average temperature of 14 Celsius degrees, the jump can be improved by around 8 cm, compared to that one performed at the see level. It is to be mentioned that the jumper's mass is pretty important, as one can be seen from the next graph.

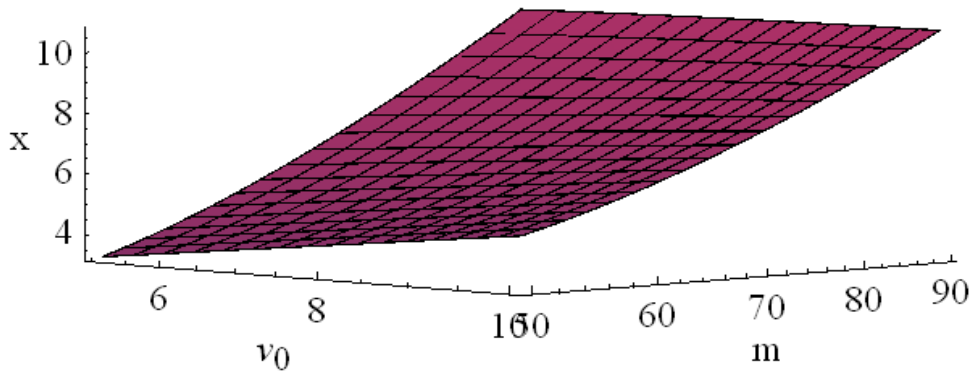

Fig.2

Adjusting the values of $y_0$, $\rho$, $m$, $v_0$, $c_{dx}$, $c_{dy}$, $A_x$, $A_y$, $x_0$ and y, this methodology can be applied at all throwing sports like hammer, shot put, discus, javelin.